\documentclass[aps,prb,twocolumn,superscriptaddress,showpacs]{revtex4}

\usepackage{graphicx}
\usepackage[usenames]{color}

\def\Csixty{C$_{60}$}

\def\Csixtytwo{C$_{60}^{2-}$}
\def\Csixtythree{C$_{60}^{3-}$}
\def\Csixtyfour{C$_{60}^{4-}$}

\def\Atwo{A$_2$C$_{60}$}
\def\Athree{A$_3$C$_{60}$}
\def\Afour{A$_4$C$_{60}$}
\def\Asix{A$_6$C$_{60}$}
\def\Natwo{Na$_2$C$_{60}$}
\def\Nathree{Na$_3$C$_{60}$}
\def\Nafour{Na$_4$C$_{60}$}
\def\Kone{KC$_{60}$}
\def\Kthree{K$_3$C$_{60}$}
\def\Kfour{K$_4$C$_{60}$}
\def\Rbfour{Rb$_4$C$_{60}$}
\def\Csone{CsC$_{60}$}
\def\Csfour{Cs$_4$C$_{60}$}
\def\cm-1{cm$^{-1}$}
\def\T1u{$T_{1u}$}
\def\Dhd{$D_{3d}$}
\def\Dod{$D_{5d}$}
\def\D2h{$D_{2h}$}

\begin{document}

\title{Phase segregation on the nanoscale in Na$_2$C$_{60}$}
\author{G. Klupp}
\email{klupp@szfki.hu}
\homepage{www.szfki.hu/~klupp}
\author{P. Matus}
\affiliation{Research Institute for Solid State Physics and Optics, Hungarian Academy of
Sciences, P.O. Box 49, Budapest, Hungary H-1525}
\author{D. Quintavalle}
\affiliation{Institute of Physics, Budapest University of
Technology and Economics, P.O. Box 91, Budapest, Hungary H-1521}
\affiliation{Solids in Magnetic Fields Research Group of the
Hungarian Academy of Sciences, P.O. Box 91, Budapest, Hungary
H-1521}
\author{L.F. Kiss}
\author{\'E. Kov\'ats}
\affiliation{Research Institute for Solid State Physics and Optics, Hungarian Academy of
Sciences, P.O. Box 49, Budapest, Hungary H-1525}
\author{N.M. Nemes}
\affiliation{Instituto de Ciencia de Materiales de Madrid (CSIC),
Cantoblanco, E-28049 Madrid, Spain}
\author{K. Kamar\'as}
\author{S. Pekker}
\affiliation{Research Institute for Solid State Physics and Optics, Hungarian Academy of
Sciences, P.O. Box 49, Budapest, Hungary H-1525}
\author{A. J\'anossy}
\affiliation{Institute of Physics, Budapest University of
Technology and Economics, P.O. Box 91, Budapest, Hungary H-1521}
\affiliation{Solids in Magnetic Fields Research Group of the
Hungarian Academy of Sciences, P.O. Box 91, Budapest, Hungary
H-1521}
\date{\today}

\begin{abstract}
\Natwo\ is believed to be an electron-hole counterpart of the Mott-Jahn--Teller insulator \Afour\ salts. We
present a study of infrared, ESR, NMR spectroscopy, X-ray diffraction, chemical composition and neutron
scattering on this compound. Our spectroscopic results at room temperature can be reconciled in a picture of
segregated regions of the size 3--10 nm. We observe a significant insulating \Csixty\ phase and at least two more
phases, one of which we assign to metallic \Nathree. The separation disappears on heating by jump diffusion of
the sodium ions, which we followed by neutron scattering. Above $\sim$460 K we see infrared spectroscopic
evidence of a Jahn--Teller distorted \Csixtytwo\ anion.

\end{abstract}


\maketitle

\section{Introduction}

The fullerene molecule, C$_{60}$, has three degenerate lowest unoccupied molecular orbitals (LUMO) and is a good
electron acceptor that forms mono- through hexaanions relatively easily. The physical properties of the salts
formed by the fulleride anions with alkali cations are in sharp contrast with a simplistic view of rigid-band
filling. Solid \Csixty\ and \Asix\ (with A=K, Rb) both have cubic structures and are non-magnetic band insulators
with an empty and full band, respectively, derived from the LUMO orbitals.\cite{erwin92} A rigid-band model would
suggest that all materials with partial filling (and the simplest possible cubic structure with one molecule per
unit cell) must be either metals or magnetic Mott insulators.\cite{lu94} However, this is not the case.
A$_{3}$C$_{60}$ salts have a half-filled band and are metals that are superconducting at low
temperature,\cite{hebard91} but none of the other compositions have a metallic ground state. Although the
low-temperature properties of cubic AC$_{60}$\ salts are difficult to establish (C$_{60}^-$ ions polymerize below
400~K\cite{steph94,pekker94}), the magnetic susceptibility of the monomeric forms at high temperature is
characteristic of paramagnetic Mott insulators.\cite{bommeli95} The nonmagnetic insulators \Afour\ (A = K,Rb,Cs)
are the most thoroughly investigated series in this family. It was for these systems that the concept of the
Mott--Jahn--Teller nonmagnetic insulator\cite{fab97} has been put forward.

The argument combining Mott localization and Jahn--Teller effect
can be summarized as follows. In order to minimize molecular
energy, C$_{60}^{n-}$ ions undergo Jahn--Teller distortion and
subsequent splitting of the LUMO levels. The Jahn--Teller
splitting is usually much smaller than the bandwidth in a
conventional solid, but Mott localization can lead to band
narrowing and singlet ground states for evenly charged ions.
Singlet-triplet excitations in such a model correspond to a small
spin gap ($\approx$~0.05~eV), \cite{brouet02a} whereas charge
excitations from one molecule to the next would cause a transport
and optical gap of 0.5 eV.\cite{knupfer97} The latter corresponds
to the transition between Jahn--Teller split levels of $t_{1u}$
origin, which would be optically forbidden on the same molecule.

Molecular calculations\cite{chancey97} show that the most significant Jahn--Teller distortion occurs for evenly
charged ions \Csixtytwo\ and \Csixtyfour\, and involves elongation and flattening, respectively, along a three-
or fivefold axis, leading to a fulleride ion with \Dhd\ or \Dod\ point group. External fields, e.g., the Coulomb
potential of the cations, can stabilize a distortion along a twofold axis, resulting in \D2h\ symmetry. This is
the case in orthorhombic \Csfour. \cite{dahlke98,dahlke02} In \Kfour\ and \Rbfour\ with average tetragonal
structure\cite{kuntscher97} a frustration exists between molecular and crystal symmetry. In these salts, there is
experimental evidence for a Mott--Jahn--Teller insulator ground state with low-lying excited
states\cite{knupfer97,brouet02a} and for the dominant role of molecular degrees of freedom over crystal field at
higher temperatures in vibrational spectra.\cite{kamaras02, klupp06}

The electronic structure of \Atwo\ salts is related to that of A$_{4}$C$_{60}$ by electron-hole symmetry and thus
they should also be insulators with the Jahn--Teller distortion in C$_{60}^{2-}$ ions lifting the degeneracy.
Direct comparison is hampered by the fact that of the \Atwo\ family only \Natwo\ has been prepared so far, and
its natural electron-hole counterpart, \Nafour, forms a polymer at room temperature.\cite{oszi97} \Natwo,
nevertheless, is considered as an ideal material to test the role of Jahn--Teller distortions since in addition
to the partial filling with an even number of electrons it crystallizes in a face-centered cubic structure above
319~K,\cite{yildirim94, yildirim93} where the threefold degeneration of both t$_{1u}$ LUMO's and T$_{1u}$
vibrational transitions remains unchanged. \Natwo\ has indeed been investigated by several authors
\cite{brouet02a,kubozono99,petit94} but findings and interpretations by different groups have been contradictory.
A detailed NMR and ESR study\cite{brouet02a} interpreted the temperature dependence of the spin susceptibility in
the framework of a Mott--Jahn--Teller singlet ground state with low-energy triplet excitations. The authors
concluded that the similar behavior of \emph{bct} \Kfour\ and simple cubic \Natwo\ proves the dominant role of
molecular distortions over that of the environment. The spin-lattice relaxation at low temperature indicated that
the band degeneracy is not completely lifted and the ground state might be weakly metallic. Kubozono \emph{et
al.}\cite{kubozono99} interpret a sudden change at 50 K in the susceptibility and the Raman spectrum as a sign of
a metal-insulator transition. Reference \onlinecite{petit94} reports hysteretic behavior in the ESR spectra in
the temperature range 300--400~K.

In this paper, we suggest that Na$_{2}$C$_{60}$ is not a simple electron-hole analogue of \Csixtyfour, but it is
yet another example of inhomogeneous charge distribution on the nanoscale. Spatial inhomogeneity is not uncommon
in fulleride salts. Metastable cubic \Csone\ seems to be a non-magnetic insulator with a charge
disproportionation into C$_{60}^{2-}$ ions and \Csixty.\cite{brouet99,brouet02b} Segregation of potassium ions
into \Kthree\ within a continuous \Csixty\ lattice has been observed for \Kone.\cite{faigel95} Disproportionation
may also play a role in \Athree: a dynamic charge fluctuation of C$_{60}^{3-}$ ions into \Csixtytwo\ and
\Csixtyfour\ would stabilize the metallic character (Ref. \onlinecite{brouet01}). $^{23}$Na NMR and $^{13}$C
magic-angle-spinning NMR data\cite{schurko04} in nominally \Nathree\ samples also suggest formation of distinct
regions of fulleride ions with various charge states. In \Natwo, we present evidence from infrared, ESR and NMR
spectroscopy and neutron scattering that Na$^+$ ions are inhomogeneously distributed on the scale of a few
lattice constants within the ordered \Csixty\ lattice. This inhomogeneous distribution leads to at least two
phases, one of which is \Csixty\ and one metallic \Nathree. The material appears homogeneous above $\sim$460~K
where Na diffusion is fast.

Single phase alkali fulleride salts are difficult to prepare and differences in experimental findings often arise
from variations in composition. However, for \Natwo\ most groups report the same X-ray diffraction pattern. A
well-defined orientational transition is observed at 319~K;\cite{yildirim94, yildirim93} at room temperature and
below the \Csixty\ molecules are ordered in a $Pa\overline{3}$ structure. The multiphase structure of \Natwo\
resolves the controversy between results of different groups: this controversy stems more from the varying
sensitivity of specific experimental methods to the presence of distinct phases than from variations in sample
preparation.

\section{Experimental}

Na$_2$C$_{60}$ was obtained by solid-state synthesis in inert gas atmosphere in a dry box. Stoichiometric amounts
of Na metal and C$_{60}$ powder were heated in a stainless steel capsule. The typical annealing sequence was
first 23~days at 350$^{\circ}$C followed by 7 days at 450$^{\circ}$C. To homogenize the samples we reground them
about once every five days. The progress of the reaction was monitored by X-ray diffraction (XRD) and
room-temperature infrared spectroscopy. After the final step, XRD results were identical to those in the
literature.\cite{yildirim93} Within the spatial resolution of XRD, a \textit{single} $Pa\overline{3}$ phase with
a lattice constant of 14.18~$\mbox{\AA}$ was found. At this point we tested the high-temperature infrared
spectrum as well, to prove that there is no residual \Csixty\ present. The samples were handled in inert
atmosphere (He, Ar or vacuum) throughout all the following preparation and measurement procedures.

Infrared spectra were recorded on pressed KBr pellets in a Bruker IFS 28 FTIR instrument either in a cryostat
under dynamic vacuum, or in argon atmosphere in a sealed sample holder with KBr windows, with 1 or 2~cm$^{-1}$
spectral resolution. Samples for investigation by XRD were enclosed in glass tubes and those for NMR, ESR and
SQUID in quartz tubes.

The static susceptibility was measured by a Quantum Design MPMS-5S
SQUID magnetometer between 5 and 600~K. We extracted the
susceptibility from fixed-temperature magnetization data between 1
and 5~T. The low-field part was omitted in order to get rid of
spurious ferromagnetic contributions.

High-frequency ESR spectra were measured on about 20 mg of
Na$_2$C$_{60}$ powder in a home-built spectrometer operating at
225~GHz (corresponding to a resonance magnetic field of 8.1~T).
The 9~GHz spectrum was recorded with a commercial Bruker ELEXSYS
500 spectrometer.
$\alpha,\gamma$-bisdiphenylene-$\beta$-phenylallyl (BDPA) was used
as standard g-factor reference (g = 2.0025).

NMR spectra were obtained by Fourier transformation of free
induction decay (FID) signals. The FID signals were collected by a
Tecmag Apollo HF spectrometer in 8.9~T static magnetic field
(corresponding $^{13}$C frequency: 95~MHz). The frequency scales
used are fractional frequency shifts  $\Delta f/f$ from the
standard reference solution of tetramethylsilane (TMS) with
positive shifts indicating higher resonant frequencies.

For neutron scattering measurements, 1.2~g of Na$_2$C$_{60}$\ powder was placed in an annular aluminum sample
holder. Temperature-dependent elastic fixed-window scans (EFWS) were taken on both the sample and the empty
sample container on the High Flux Backscattering Spectrometer of the NIST Center for Neutron
Scattering.\cite{HFBS} The individual detector counts were normalized by the neutron monitor of the direct beam
and binned into 5~K intervals to improve the signal-to-noise ratio. Background correction consisted of
subtracting data measured on the empty sample container under similar conditions. These corrections never
exceeded one third of the sample counts.

\begin{figure*}
\includegraphics[]{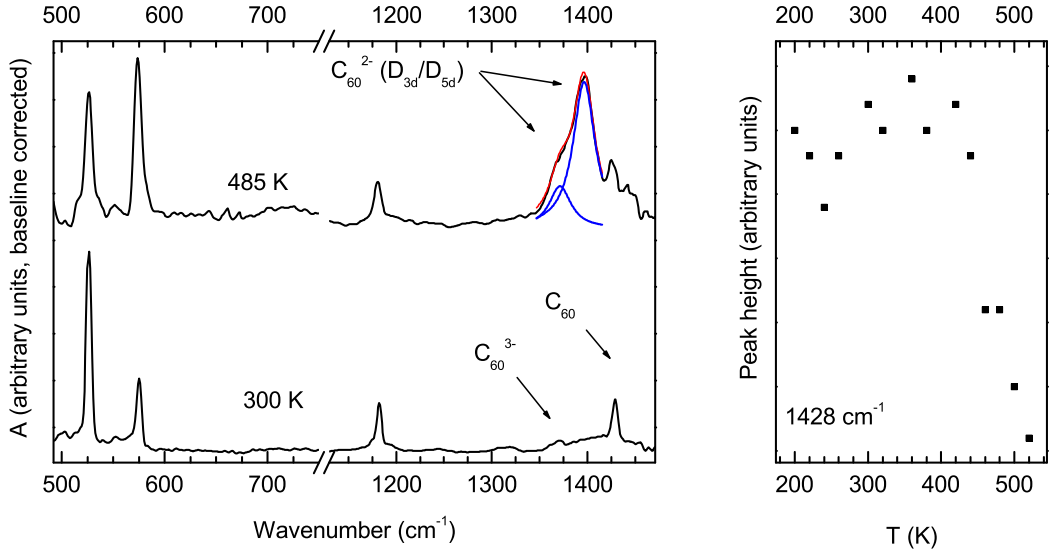}
\caption{(Color online) Left panel: infrared absorption of Na$_2$C$_{60}$ at 300 K and 485 K in the region of the
principal C$_{60}$ vibrations. Right panel: temperature dependence of the intensity of the 1428~cm$^{-1}$ line,
characteristic of neutral C$_{60}$.} \label{fig:irTdep}
\end{figure*}

\section{Results}

We investigated the samples of nominal composition \Natwo\ by infrared, ESR and NMR  spectroscopy, direct
measurement of chemical composition and neutron scattering. First we summarize the limits of quantitative
determination using the methods we combine in this study. Infrared spectra yield the vibrational frequencies of
individual fulleride anions, and reflect their charge and symmetry. While the intensity of a specific absorption
peak assigned to a species (e.g., a molecular ion with a given charge) scales linearly with its concentration,
the cross sections can differ considerably between species. In metallic systems, the free-electron background
further complicates the spectra by smearing the vibrational peaks. ESR gives information on species with unpaired
spins. The high-frequency resonances of various phases are well resolved and the relative spin susceptibilities
can be determined. NMR intensities scale with the amount of the nucleus investigated, but in C$_{60}^{n-}$
anions, the position of the peaks does not shift enough with charge to be resolved.

Taking into account all the limitations and combining the information from individual methods, we conclude that
none of the results at room temperature is compatible with the picture of a homogeneous material containing only
C$_{60}^{2-}$ molecular ions. Below we describe the findings by individual methods in detail.

\subsection{Infrared spectroscopy}

Figure \ref{fig:irTdep} shows the infrared spectra at 300~K and 485~K. There is an obvious difference between the
two spectra, most remarkable in the range of the highest-frequency T$_{1u}$(4) mode. On cooling back from 480~K
to room temperature, the original 300~K spectrum reappears after about two weeks.

We discuss first the high-temperature spectrum which comfirms the assumption of a homogeneous \Natwo\ phase and
then the low-temperature phase which contradicts this assumption.

Of the four infrared-active vibrations of C$_{60}$, the highest frequency \T1u\ mode is most sensitive to charge
$n$ when C$_{60}^{n-}$\ anions are formed. From the empirical linear relationship between the T$_{1u}$(4)
frequency and $n$,\cite{pichler94} we estimate the frequency in C$_{60}^{2-}$ to be about 1380~\cm-1, possibly
with a splitting due to the molecular Jahn--Teller effect.\cite{long98,kamaras02} Indeed, above 460~K two bands
are resolved at 1369 and 1394 cm$^{-1}$, close to the expected position of the C$_{60}^{2-}$ absorption. The
twofold splitting of the mode indicates $D_{3d}$ or $D_{5d}$ molecular symmetry, i.e., a Jahn--Teller distortion
of \emph{molecular} origin.\cite{chancey97}

The room-temperature spectrum (Fig.~\ref{fig:irTdep}) shows a strong peak at 1428~\cm-1, typical of neutral
C$_{60}$, and a smeared one with a weak peak at 1370~\cm-1.\ The spectrum does not change on cooling between
300~K and 80~K. The low-temperature spectra are in contradiction with expectations for a homogeneous \Natwo\
compound. The weak peak at 1370 cm$^{-1}$ is at the position expected for \Nathree. A$_3$C$_{60}$ salts are
metallic and show less intense vibrational peaks due to the large background from free-carrier
absorption.\cite{pichler94,martin93}

The 1428 \cm-1\ mode cannot be assigned to unreacted residual C$_{60}$\ since it disappears on heating above
460~K and reappears on cooling back in a reproducible manner. The data rather suggest that the material contains
a mixture of fulleride ions. We can positively identify neutral \Csixty\ molecules and \Csixtythree\ ions at room
temperature and below, and \Csixtytwo\ dianions at higher temperature. We cannot tell from the spectra whether
the different vibrational signatures come from isolated sites embedded in some other matrix or from highly
concentrated phases. However, from the metallic background it is probable that the phase containing C$_{60}^{3-}$
is a metal, presumably \Nathree. We cannot exclude the presence of small amounts of other fulleride anions at low
temperature, either; their vibrational peaks may be buried in the background around 1400~cm$^{-1}$.

\subsection{Electron spin resonance and static susceptibility}

\begin{figure}
\includegraphics{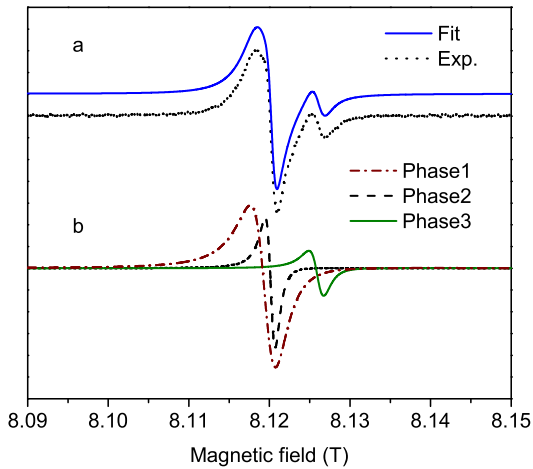}
\caption{(Color online) Decomposition of the 225 GHz ESR spectrum into lines corresponding to segregated phases
at 40 K. a) experimental spectrum and sum of three fitted lines; b) decomposition of experimental line into three
segregated phases.} \label{fig:esr40}
\end{figure}

The static susceptibility measured by SQUID magnetometry at 4~T shows three characteristic temperature ranges:
i.) T$<$100 K: a Curie-like paramagnetic increase with decreasing temperature; ii.) 100 K $<$ T $<$ 300 K: an
approximately constant susceptibility with $\chi=1.8\cdot 10^{-7}$ emu/gOe; and iii.) 300 K $<$ T $<$ 600 K: a
continuous increase to about $6\cdot 10^{-7}$ emu/gOe. The zero value of the static susceptibility is not
corrected for core electron effects. The temperature-dependent part of the susceptibility measured by ESR at
9~GHz in Ref.~\onlinecite{brouet02a} shows the same three characteristic ranges.

\begin{figure}
\includegraphics{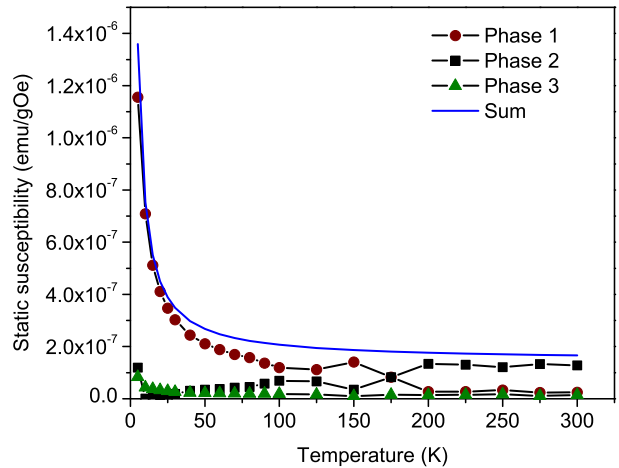}
\caption{(Color online) Decomposition of static spin susceptibility into various phases from ESR. The absolute
values of the susceptibility were determined from the static (SQUID) measurements.} \label{fig:esrfullT}
\end{figure}

In the 225~GHz high-frequency ESR spectra, lines of three phases were resolved at all temperatures between 5 and
450~K. In a conventional spectrometer at 9~GHz, these lines were not resolved; a single line of 9~G peak-to-peak
derivative linewidth was found at 300~K. A 225~GHz spectrum taken at 40~K is shown as an example in
Fig.~\ref{fig:esr40}. We attribute the three lines to three different segregated phases in the nominally
Na$_{2}$C$_{60}$ material. The ESR line intensities are proportional to the spin susceptibilities of the various
phases and are shown in Fig.~\ref{fig:esrfullT}. The full ESR spectra were decomposed into
three Lorentzian lines and the fractional intensities for each line $i$, $%
I_{i}$ $=\chi _{i}V_{i}/C$, (the spin susceptibility, $\chi _{i},$ times the
fractional volume,\ $V_{i}$) were determined. The temperature dependence of
the full ESR intensity, $C=\Sigma I_{i}$, was not measured directly,
instead, we normalized the total ESR intensity data at each temperature to
the static susceptibilities measured by SQUID.

At low temperature (Fig.~\ref{fig:esr40}), the static susceptibility is dominated by a single phase (Phase~1)
with a Curie-like susceptibility corresponding to 1.3\% of C$_{60}^-$ (or other spin 1/2) free radicals with
respect to the entire \Csixty\ content of the sample. We denote by Phase~2 the phase with the largest ESR
intensity at ambient temperature. Phase~3 has a smaller g-factor than Phases 1 and 2 and could be followed in a
wide temperature range despite the small intensity. Since the corresponding susceptibility does not vary
strongly, it is possible that Phase~3 is metallic and may be assigned to the \Nathree\ observed by infrared
spectroscopy.

\begin{figure}
\includegraphics{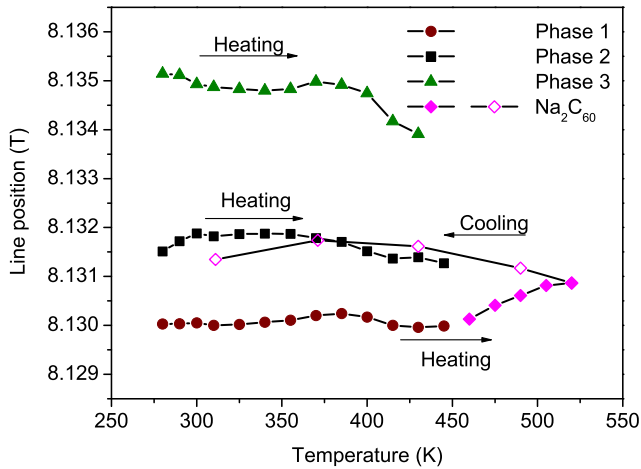}
\caption{(Color online) Temperature dependence of 225 GHz ESR line positions of nominal \Natwo\ (\emph{g} factor
of Phase~1 at 300~K: 2.0012). Full (empty) symbols refer to heating (cooling). The three lines corresponding to
various segregated phases merge into a single line on heating above 450~K indicating an increased homogeneity due
to Na$^+$ diffusion. On cooling to 300~K, a metastable phase is established that disappears after a few days.}
\label{fig:esrht}
\end{figure}

At about 450~K a structural transition takes place, the three low-temperature phases disappear and are replaced
by a new phase (Fig.~\ref{fig:esrht}). This is the same temperature region where the 1428 cm$^{-1}$ infrared line
decreases. The single Lorentzian line indicates an increased homogeneity. The material is, however, still
somewhat inhomogeneous between 450 and 500 K since spectra on heating and cooling are different. On rapid cooling
(2.5 K/min) the sample is quenched into a homogeneous metastable state. The original phase-segregated material
reappears after about two weeks at room temperature.

The main conclusion to be drawn from the ESR spectra is that in equilibrium the sample is inhomogeneous below
450~K. C$_{60}^{n-}$ paramagnetic ions with different \emph{g} factors due to differing environments will show a
single resonance at an average \emph{g} factor if exchange interaction energies between them are larger than
$(\Delta g/g)\hbar \omega _{L}$ where $\Delta g$ is the difference in \emph{g} factors and $\omega _{L}$ is the
Larmor frequency. Fulleride metals or semiconductors with localized C$_{60}^{n-}$ ions therefore exhibit a single
line from the conduction and localized electrons. Neighboring paramagnetic C$_{60}^{n-}$ ions will not show
separate resonances even if they are embedded in an insulator. Thus paramagnetic C$_{60}^{n-}$ ions in Phase~1
must be isolated from Phase~2 by nonmagnetic C$_{60}^{n-}$ ions. The simplest assignment is that Phase~1 is solid
neutral \Csixty\ doped with a low concentration of isolated Na$^+$ ions or small groups of Na$^+$ ions. Phase~1
must be a non-negligible fraction of the material since it contains over 1\% isolated free paramagnetic
C$_{60}^{n-}$ ions surrounded by a shell of nonmagnetic ions. We believe this phase is the neutral \Csixty\ phase
detected by $^{13}$C NMR containing a small concentration of Na$^+$ ions and ESR active C$_{60}^{-}$.  A mixture
of Phase~2 and Phase 3 appears to be the phase studied in detail by $^{13}$C and $^{23}$Na NMR and attributed to
\Natwo\ by Brouet \emph{et al.}\cite{brouet02a} Above 450 K the Na$^{+}$ ions diffuse rapidly in the crystal
lattice and the sample becomes homogeneous.

\subsection{Nuclear magnetic resonance}

\begin{figure}
\includegraphics{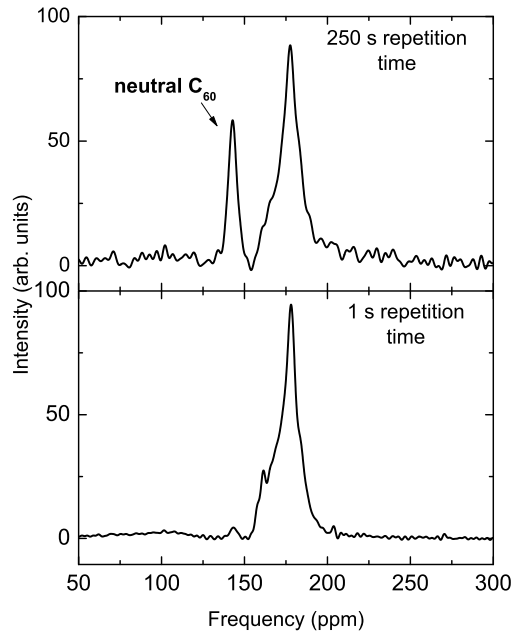}
\caption{NMR spectra at 300 K with two different repetition rates.
At fast repetition rates only \Csixty$^{n-}$ ions with short
spin-lattice relaxation time are recorded. At slow repetition
rates the spectrum of neutral \Csixty\ is also observed. }
\label{fig:NMR}
\end{figure}

Infrared absorption measurements detected C$_{60}^{n-}$ molecular ions with various charge states below 460~K and
high-frequency ESR experiments revealed several phases in nominal \Natwo. The C$_{60}$ concentration can be
determined using NMR spectroscopy. Figure~\ref{fig:NMR} shows the $^{13}$C NMR spectrum of a powder sample at
300~K. With conventional repetition rates of about 1~$\mathrm{s}^{-1}$ we observe a single NMR line around
178~ppm with chemical-shift anisotropy (shown in the lower panel of Fig.~\ref{fig:NMR}). This spectrum is very
similar to that already published by Rachdi \emph{et al.}\cite{rachdi97} With a much longer repetition time
(250~seconds), however, a further line appears at 143~ppm (upper panel of Fig.~\ref{fig:NMR}), which we assign to
neutral \Csixty.\cite{tycko91} As expected for pure \Csixty, the $^{13}$C spin-lattice relaxation time is long,
about 50~seconds at 300~K. We used a repetition time more than five times the $T_1$ spin-lattice relaxation time
of this slowly relaxing spectral component to determine the spectral weights and estimate 25$\pm$5\% neutral
\Csixty\ content in the material. This high value is in agreement with the determination of chemical composition
(see below).

Unlike the three phases in high-frequency ESR spectra, only two are resolved by NMR. We assign these as
containing neutral and charged C$_{60}$, respectively, where the charged part contains a mixture of phases,
including both \Nathree\ and Phase 2. The $^{13}$C spin-lattice relaxation of charged C$_{60}^{n-}$ is many
orders of magnitude faster than that of \Csixty\ ($T_1$ roughly 100 ms compared to 100~s), because of the
hyperfine interaction with conducting electrons. The large difference in relaxation rates indicates that
cross-coupling between nuclei in the charged and neutral phases is small, so these nuclei must be well separated
in space. More refined methods, e.g., magic angle spinning NMR, would probably be able to resolve more phases.

\begin{figure*}
\includegraphics{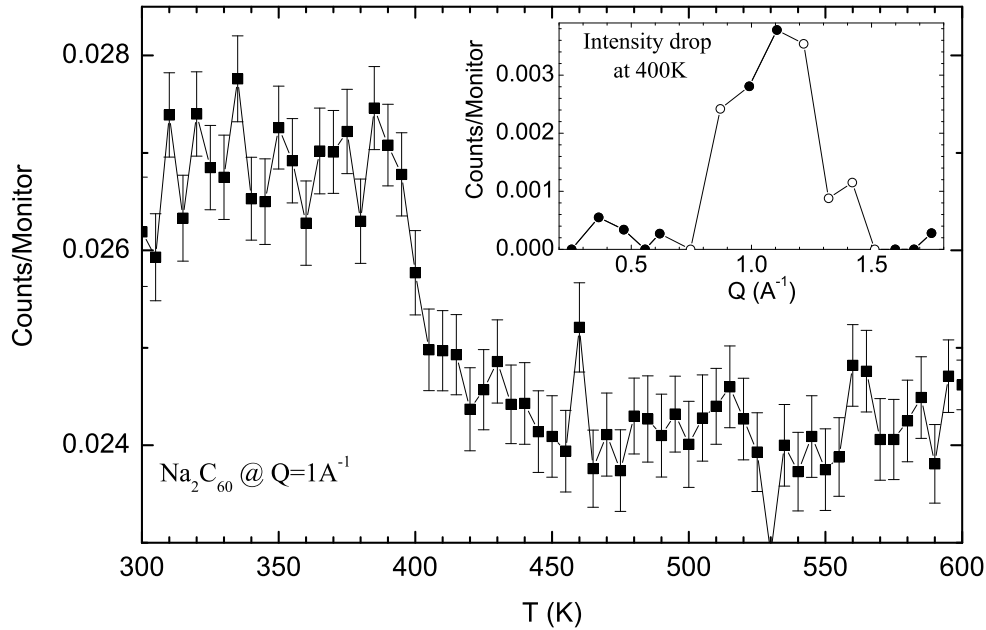}
\caption{Elastic fixed-window scan (EFWS) intensity of Na$_2$C$_{60}$ at $Q=1~\mbox{\AA}^{-1}$. Inset: \emph{Q}
dependence of the drop in EFWS intensity around 400~K (with empty circles indicating detectors obscured by
Na$_2$C$_{60}$ diffraction peaks).} \label{fig:HFBS}
\end{figure*}

\subsection{Chemical composition}

In usual multiphase fulleride samples the simplest method to separate pristine \Csixty\ from its salts is
extraction by toluene.\cite{carrard96} We extracted C$_{60}$ from the nominally Na$_2$C$_{60}$ material by
soaking it in toluene for 11 days. Following this treatment, we determined the \Csixty\ content of the solution
by high-pressure liquid chromatography (HPLC). The weight percentage of \Csixty\ can be calculated from the
concentration of the \Csixty\ solution and the weight of the starting material. To obtain the mole percentage of
C$_{60}$ we assume that the material contains only one Na$_x$C$_{60}$ phase besides the C$_{60}$ phase, but we do
not assume that the total stoichiometry of the material is nominal Na$_2$C$_{60}$. If the Na-fulleride phase in
the material were Na\Csixty\ then the \Csixty\ content would be 27 mole percent. The other extreme,
Na$_{12}$C$_{60}$, would give 33~mole percent. This estimated composition of  27-33\% \Csixty\ is in reasonable
agreement with the NMR results.

\subsection{Neutron scattering}

Diffusion of sodium was directly observed using neutron scattering. In this measurement, the incident neutron
energy is fixed at 2.08~meV and scattering processes are detected near the elastic line within the energy window
of the resolution of the instrument (1~$\mu$eV). Thus when a dynamic process, such as sodium jump diffusion,
becomes faster than the corresponding timescale of 0.8~ns, the measured EFWS intensity decreases.\cite{becker03}
Figure~\ref{fig:HFBS} shows the temperature dependence of the elastic line intensity at $Q=1~\mbox{\AA}^{-1}$.
This intensity shows typical Debye-Waller type decrease overall, but an unusual drop appears around 400~K over
approximately 20~K. Our interpretation is that the jump diffusion of sodium ions between tetrahedral and
octahedral sites becomes fast enough above 400~K to be resolved by the instrument and thus the incoherent
scattering contribution of the sodium ions is removed from the EFWS intensity. The EFWS intensity of the
background is much smaller than that of \Natwo\ and its temperature dependence is featureless.

We can get an idea of the nature of the Na$^+$ motion by looking at the \emph{Q} dependence of the anomalous drop
of the EFWS intensity, shown in the inset of Fig.~\ref{fig:HFBS}. The largest drop of the elastic intensity
occurs near Q=1.1~$\mbox{\AA}^{-1}$, which in a jump-diffusion model corresponds to a jump distance of
4.0$\pm$0.5~$\mbox{\AA}$.\cite{dewall02} The \emph{Q} resolution is determined by the fact that the detectors of
the spectrometer integrate over 0.2~$\mbox{\AA}^{-1}$ regions in \emph{Q}. Diffraction peaks of
Na$_2$C$_{60}$\cite{rosseinsky92} between 0.5~$\mbox{\AA}^{-1}$ and 0.9~$\mbox{\AA}^{-1}$ and also between
1.2~$\mbox{\AA}^{-1}$ and 1.6~$\mbox{\AA}^{-1}$ obscure the EFWS, as indicated by empty circles, so these points
are less reliable and we cannot make further inferences from these data about the nature of the Na$^+$ jumps.

Due to steric hindrance, the diffusion of Na$^+$ ions can only proceed through the trigonal point between the
tetrahedral and octahedral voids, i.e., in the $\langle$111$\rangle$ direction. In this direction, removed
4.0~$\mbox{\AA}$ from the tetrahedral site, there is an off-centered octahedral site. This site is special
because it has the same nearest-neighbor distances as the tetrahedral site, so that the Coulomb interaction
between the Na$^+$ cation and the fulleride anions remains optimal. Occupation of low-symmetry octahedral
positions by Na$^+$ ions was previously suggested by Schurko \emph{et~al.},\cite{schurko04} based on $^{23}$Na
NMR experiments on samples with nominal composition \Nathree.

\section{Discussion}

All the results presented above can be reconciled if we assume a
synproportion/disproportion reaction
\[3 \mbox{C}_{60}^{2-} \rightleftharpoons\mbox {C}_{60} + 2 \mbox{C}_{60}^{3-}\]
taking place on cooling after preparation from left to right and
on heating from right to left.  Combining the data, we can also
draw quantitative conclusions as to the stoichiometry, the size of
the domains, and the time scale of the dynamics.

\begin{figure}
\includegraphics[scale=0.1]{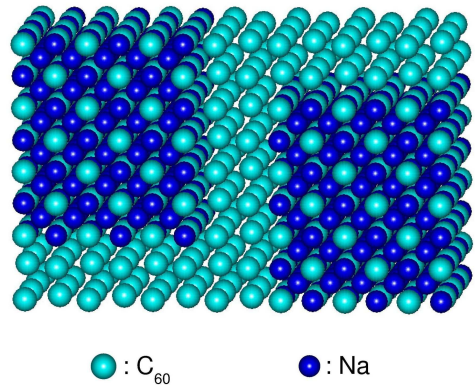}
\caption{(Color online) Schematic view of segregated regions with and without sodium ions. Estimated area of the
regions is between 3 and 10~nm.} \label{fig:plot}
\end{figure}

ESR, infrared and NMR measurements all confirm the presence of neutral \Csixty\ in the material. While the former
two methods yield qualitative results, NMR also provides quantitative data. From NMR and chemical composition
studies, we obtain the range for the C$_{60}$ content to be between 20 and 33 mole percent at room temperature.
We illustrate an idealized picture of the fully nanosegregated phases in Fig.~\ref{fig:plot}, consisting of 33\%
\Csixty\ and 67\% \Nathree. In reality at finite temperatures the reaction is probably not complete, partly
because of entropy reasons and partly because the interfaces between the segregated regions must consist of
fulleride ions only partially surrounded by sodium ions. These interfaces may be responsible for the additional
ESR signal, the broad NMR line and the broad background of the infrared spectrum between 1350 and 1440 cm$^{-1}$.

The spectroscopic results apparently contradict the X-ray diffraction data which see the material as a single
phase. We have to take into account, though, that while diffraction measures the average crystal structure, over
a relatively large area, spectroscopy is sensitive to local structure, i.e., domains of a few molecules. X-ray
diffraction averages out inhomogeneities less than about 100~$\mbox{\AA}$, especially if the structure of the
constituting phases is similar. In \Nathree, the fulleride ions occupy $fcc$ sites identical to those of \Csixty\
molecules in a \Csixty\ crystal, and the lattice constant of \Nathree\ of 14.19~$\mbox{\AA}$ is very close to
that of \Csixty, 14.15~$\mbox{\AA}$.

The lower size limit of the homogeneous domains can be estimated from the metallic nature of the \Nathree\ phase.
For metallic behavior, a thickness of a few molecules is required. The lattice constants are about 1.4 nm and the
size of homogeneous domains has to be in the range of 3 to 10 nanometers.

At high temperature the nanosegregated phase changes into a homogeneous phase by sodium diffusion in a more or
less unaltered \Csixty\ lattice. We observed in the neutron scattering experiments that sodium ions jump from
site to site on the time scale of 0.8~ns at 400~K. In a nanosegregated structure with mobile sodium ions,
diffusion leads to homogenization. The extra electrons migrate along with the ions because of Coulomb attraction,
eventually leading to charge equilibrium in the system. Once the diffusion rate is high enough to achieve this
equlibrium, the diffusion does not cause further charge migration and the material appears homogeneous by all
spectroscopic methods. ESR shows a hysteretic phase segregation: on heating we observe segregated phases up to
450~K. Above 450 K there are no well separated conducting and insulating regions and the resonance is narrowed
into a single line by electronic spin diffusion. This close-to-homogeneous phase is preserved for some time on
cooling. A metastable phase with a single ESR line is observed on cooling rapidly from 520~K to 300~K. Infrared
spectra also show homogeneous charge distribution above 460~K, with indications of a metastable phase surviving
on the time scale of weeks after cooling to room temperature.

The twofold splitting of the T$_{1u}$(4) infrared line at high temperature suggests that the symmetry of the
molecules in the homogeneous phase is \Dhd\ or \Dod. At first sight, this symmetry seems to contradict the cubic
crystal structure, but can be explained as follows. Since in the high-temperature phase of \Natwo\ the molecules
show quasifree rotation,\cite{yildirim94, yildirim93} the distorting effect of the crystal field is averaged out.
The absence of crystal field, however, does not mean that the shape of the molecule will stay spherical; it will
be determined by the interaction of molecular vibrations with the two extra electrons occupying the LUMO (the
molecular Jahn--Teller interaction).\cite{chancey97} In case of dianions, the result is a set of equivalent \Dhd\
or \Dod\ distortions in different (symmetry-related) directions. These equivalent distortions transform into each
other by a special motion called \emph{pseudorotation},\cite{chancey97} and their average will be icosahedral.
(Tomita \emph{et al.}\cite{tomita05} present a detailed explanation of the effect and provide experimental
evidence from near-infrared spectroscopy of isolated C$_{60}^-$ ions.) Nevertheless, if the time scale of a
measurement is smaller than that of the pseudorotation, the distorted structure will be detected. This must be
the case of the infrared spectra. The high-temperature phase of \Natwo\ is another example of the dynamical
Jahn--Teller effect in fulleride salts. The resulting symmetry strongly resembles that occurring in the
high-temperature phase of \Afour\ compounds,\cite{klupp06} proving the argument by Brouet \emph{et al.} about the
molecular interactions dominating over crystal field effects in both systems.\cite{brouet02a}

\section{Conclusions}

We have presented ample evidence from a wide range of analytical and spectroscopic methods that the fulleride
salt with nominal composition \Natwo\ at room temperature forms nanosegregated domains with a large portion of
\Csixty\ (containing Na$^+$ ions in a low concentration) and other phases with high Na$^+$ ion concentration,
including metallic \Nathree. We estimate the amount of \Csixty\ in the material to be 20-33\% at 300~K. A
synproportion reaction takes place on warming, homogeneity is achieved by the diffusion of sodium ions. In the
homogeneous material, the \Csixtytwo\ anions undergo dynamic Jahn--Teller distortion to \Dhd/\Dod\ symmetry.

Our samples showed the same X-ray diffraction pattern, 9 GHz ESR
spectrum and low retention time NMR spectrum as those investigated
previously. We expanded these experiments to different measuring
conditions and employed additional methods such as infrared
spectroscopy, chemical composition determination and neutron
scattering. All these experiments can be reconciled by assuming
the nanosegregation model at low temperature and homogeneous
\Natwo\ at high temperature. The results clarify the main reason
for the contradictory results in the literature. They also
underline the necessity of new experiments for the verification of
the Mott--Jahn--Teller picture of insulating fulleride salts.

\begin{acknowledgments}
We gratefully acknowledge the help of G. Oszl\'anyi with X-ray diffraction measurements and extensive
discussions. Work in Hungary was supported by the Hungarian National Research Fund under grant numbers OTKA
T049338, T046700, TS049881 and T043255. This work also utilized neutron research facilities of the National
Institute of Standards and Technology, U.S. Department of Commerce, supported in part by the NSF DMR-0454672.
N.N.M. is supported by the Juan de la Cierva fellowship of the Spanish Ministry of Education.

\end{acknowledgments}

\bibliography{c60ir}

\end{document}